\documentclass[aps,prl,reprint]{revtex4-1}   

\usepackage{graphicx}
\usepackage{bm}
\usepackage{epstopdf}
\usepackage[sort&compress]{natbib}
\usepackage{amsmath}  
\usepackage{color}
\begin{document}

\title[Striations in electronegative CCRF plasmas]{Experimental observation and computational analysis of striations in electronegative capacitively coupled radio-frequency plasmas}

\author{Yong-Xin Liu$^1$, Edmund Sch\"ungel$^2$, Ihor Korolov$^3$, Zolt\'an Donk\'o$^3$, You-Nian Wang$^1$, Julian Schulze$^{2,4}$} 

\affiliation{$^1$Key Laboratory of Materials Modification by Laser, Ion, and Electron Beams (Ministry of Education), School of Physics and Optoelectronic Technology, Dalian University of Technology, Dalian 116024, China}

\affiliation{$^2$Department of Physics, West Virginia University, Morgantown, West Virginia 26506-6315, USA} 

\affiliation{$^3$Institute for Solid State Physics and Optics, Wigner Research Centre for Physics, Hungarian Academy of Sciences H-1121 Budapest, Konkoly-Thege Mikl\'os str. 29-33, Hungary}

\affiliation{$^4$Institute for Electrical Engineering, Ruhr-University Bochum, Germany}


\date{\today}

\begin{abstract}
Self-organized spatial structures in the light emission from the ion-ion capacitive RF plasma of a strongly electronegative gas (CF$_4$) are observed experimentally for the first time. Their formation is analyzed and understood based on particle-based kinetic simulations. These ``striations'' are found to be generated by the resonance between the driving radio-frequency and the eigenfrequency of the ion-ion plasma (derived from an analytical model) that establishes a modulation of the electric field, the ion densities, as well as the energy gain and loss processes of electrons in the plasma. The growth of the instability is followed by the numerical simulations.



\end{abstract}


\maketitle 

Plasmas in electronegative gases exhibit complex physical and chemical kinetics \cite{eneg2,eneg3,eneg4,eneg5,eneg6,enegA,eneg8,eneg9,Sheridan1,Sheridan2,Sheridan3}. Their main constituents are typically positive and negative ions, and electrons are only present as a minor species. Such a composition leads to unique effects, e.g., the dominant mechanism of electron energy gain is typically due to the ambipolar and drift electric fields within the {\it ion-ion plasma} bulk \cite{BulkMode,Tochikubo,ELIAS1,Perrin,Boeuf,ambipolar,WangAmbipolar}, in sharp contrast with the mechanisms in (more common) electropositive (electron-ion) plasmas where the dynamics of the boundary sheaths conveys energy to the electrons.  

Being complex dynamical systems, plasmas are susceptible to various instabilities. Strong modulations of the plasma density and light emission -- termed as ``striations'' -- have extensively been studied in electropositive DC discharges \cite{DC1,DC2,DC3,DC5}, wherein ion-acoustic or ionization waves form the basics of these features. Striations also occur in electropositive inductively coupled plasmas \cite{ICP1}, plasma display panels \cite{display1}, and in plasma clouds in the ionosphere \cite{ionosphere1,ionosphere2}. In these system the appearance of striations is explained by theories based on the {\it electron kinetics}. Little is known, however, about the nature of striations in electronegative plasmas where the {\it ion kinetics} may play the dominant role: observations have been limited to DC plasmas \cite{DCeneg1,DCeneg2}, and striations have never been observed in electronegative capacitively-coupled RF (CCRF) plasmas, to our best knowledge. Here we report the observation of striations, that form in the bulk of electronegative CCRF plasmas. The experimental observations are compared with simulation data, which allow a detailed investigation of the underlying physics. 

In the experiment (described in detail in \cite{WangAmbipolar}) the plasma is produced in CF$_4$ between two parallel electrodes made of stainless steel with a diameter of 10 cm. The gap between the electrodes is $L$=1.5 cm. The bottom electrode and the chamber walls are grounded. A sinusoidal voltage of a function generator is amplified and applied to the top electrode via a matching network. The generator is also connected to a pulse delay generator that triggers in a synchronized manner an intensified charge-coupled device (ICCD) camera for Phase Resolved Optical Emission Spectroscopy (PROES) measurements. The ICCD camera is equipped with an objective lens and an interference filter to detect the spatio-temporal emission of a specifically chosen optical transition (at 585 nm) of Ne (which is admixed at 5\% as a tracer gas). From the light emission measurements that are performed in a sequence through the RF period the spatio-temporal electron impact excitation rate from the ground state into the Ne$2p_1$-state is calculated based on a collisional-radiative model (for more details see \cite{PROES,GansProes,WangAmbipolar}). The voltage waveform is measured by a high-voltage probe and is acquired with a digitizing oscilloscope. The results are presented for a driving voltage with 8 MHz frequency and 300 V amplitude and a gas pressure of 100 Pa.

Our numerical studies are based on a Particle-in-Cell code complemented with Monte Carlo treatment of collision processes (PIC/MCC) \cite{PIC1,PIC2,EAE15}. The simulation is performed with one-dimensional spatial resolution within the electrode gap and full resolution of the velocity space. 
The gas temperature is set to 350 K. 
The emission of secondary electrons from the electrode surfaces is neglected to simplify the analysis. Electrons are reflected from the electrodes with a probability of 0.2. Besides the electrons, we trace only the dominant ion species, CF$_3^+$, CF$_3^-$, F$^-$, in our code \cite{Morris}. 
The cross sections are taken from Refs. \cite{Kurihara2000,Bonham1994,GB2,Nanbu2000,Proshina,Denpoh2000}.
For more details see Ref. \cite{EAE15}. While for a symmetrical electrode configuration, such as considered here, generally no DC self-bias should appear, we allow for the development of a DC self-bias in the simulations, since the emergence of density modulations may break the symmetry of the discharge. 

\begin{figure}[t!]
\begin{center}
\includegraphics[width=0.5\textwidth]{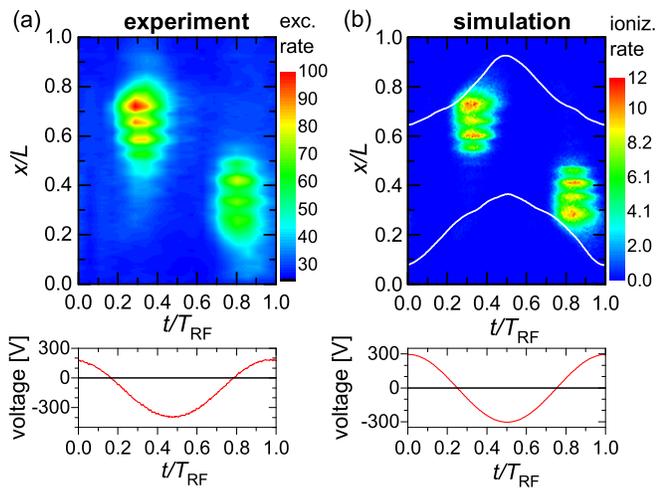}
\end{center}
\caption{(a) Measured electron-impact excitation rate and (b) simulated spatiotemporal ionization rate (with white lines showing the sheath edges). Powered electrode is at $x/L = 0$, grounded electrode is at $x/L = 1$. The panels at the bottom show the discharge voltage in the experiment and in the simulation.} 
\label{fig:simProes}
\end{figure}

Figure \ref{fig:simProes}(a) shows the experimentally determined electron-impact excitation rate spatially and temporally resolved. The excitation rate is high near the grounded electrode during the first half of the RF period and near the powered electrode during the second half. This is due to the energy gain of electrons in the electric field inside the plasma bulk at the times of the collapse of the adjacent sheaths \cite{BulkMode,Tochikubo,ELIAS1,Perrin,Boeuf,ambipolar,WangAmbipolar}. (We note that a small negative shift, i.e., a negative self-bias voltage is observable in the experiment, due to the geometrical asymmetry of the reactor, which also gives rise to a slight asymmetry of the excitation rate (cf. Fig.\ref{fig:simProes}(a)).)

The excitation rate exhibits a remarkable spatial modulation, which indicates that the number of electrons with energies above the excitation threshold of 19 eV varies periodically. In the experiment these {\it striations} are stable and observable by naked eye. These observed patterns agree qualitatively well with the layered structures found in the spatio-tempoal distribution of the ionization rate obtained from the simulation (Fig.\ref{fig:simProes}(b)). The white lines indicate the position of the edges of the sheaths adjacent to the two electrodes \cite{Brinkmann}. In the following, we discuss the physics of these striations based on the simulation data obtained for the steady state. 
  
\begin{figure}[t!]
\begin{center}
\includegraphics[width=0.45\textwidth]{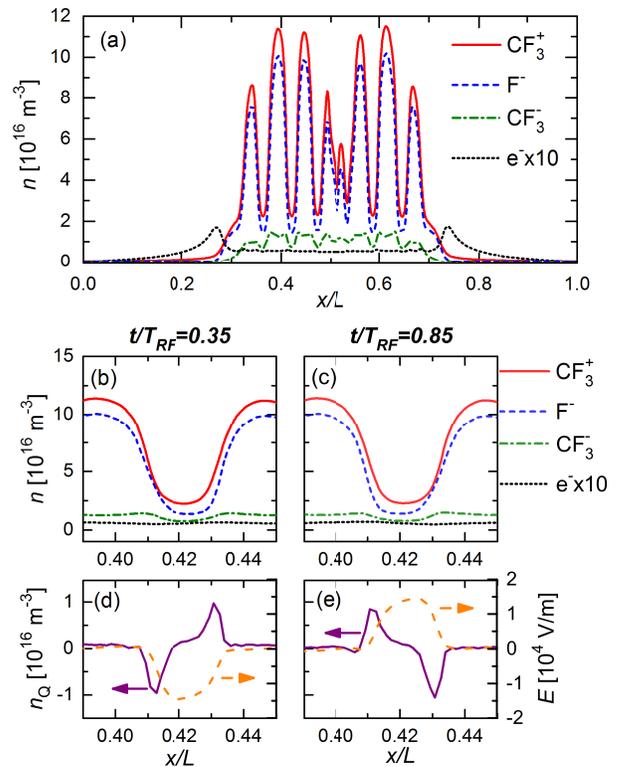}
\caption{(a) Time-averaged ion and electron density profiles obtained from the PIC/MCC simulation, (b) and (c) density profiles of charged species, (d) and (e) space charge density, $n_Q$, and electric field, $E$, at different times for 0.39-0.45 $x/L$ region. A video showing the ion motion is provided online \cite{striations_video}.} 
\label{dens}
\end{center}
\end{figure}  

The appearance of such striations is linked to strong alterations of the time averaged charged particle densities. The profiles of the CF$_3^+$, CF$_3^-$, F$^-$ ions and electrons are shown in Fig. \ref{dens}(a). Large peaks of the positive and negative ion density are found in the strongly electronegative bulk plasma region. The maxima of the striations in the excitation and ionization occur at the edges of these density peaks. They are caused by the periodic motion of the positive and negative ions into opposite directions as a consequence of their reaction to the RF modulation of the local electric field. This oscillation is possible due to their low mass as well as the low driving frequency and is illustrated in Figs. \ref{dens}(b) and (c), which show the densities in a spatial region between two neighboring striations at the times of largest ion displacement to either side, i.e., of largest space charges. The amplitude of these oscillations is small, but due to the high ion number densities and gradients significant space charge densities develop and the electric field is altered, when ions move in opposite directions (see Figs. \ref{dens}(d) and (e)). 

\begin{figure}[h!]
\begin{center}
\includegraphics[width=0.5\textwidth]{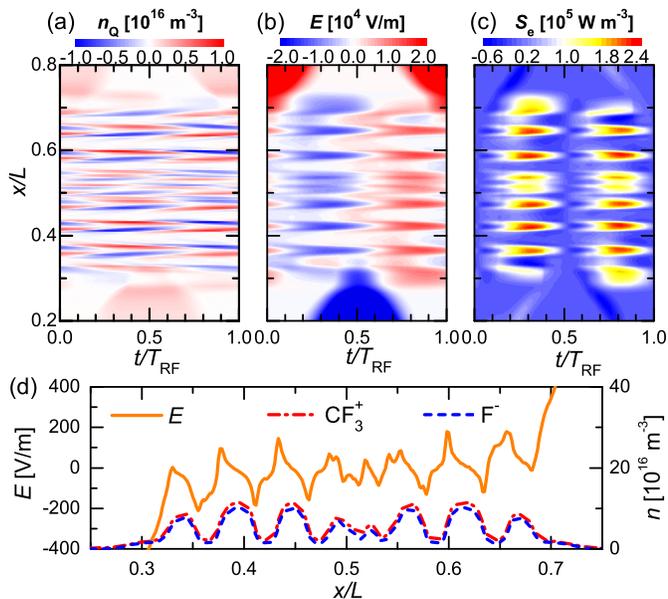}
\caption{PIC/MCC simulation results: spatio-temporal plots of the (a) space charge density, $n_Q$, (b) electric field, $E$, and (c) electron power absorption rate density, $S_e$, for the $0.2 \leq x/L \leq 0.8$ spatial region, (d) time-averaged electric field (left axis) and densities of CF$_3^{+}$ and F$^{-}$ (right axis) in the plasma bulk region.} 
\label{fig:xts}
\end{center}
\end{figure}

Figure \ref{fig:xts} shows the spatio-temporally resolved space charge, electric field, and electron power absorption in the plasma bulk. At the beginning of the RF period, a negative electric field inside the plasma bulk accelerates positive ions to the bottom (powered) electrode and negatively charged particles towards the top (grounded) electrode. The strength of this field is strongly modulated in space. Accordingly, space charges build up locally with alternating signs. Therefore, in the presence of striations and around the times of maximum positive and negative electric field in the bulk, there is a spatially alternating space charge. These space charges, in turn, generate an electric field, which is comparable in magnitude with the drift electric field generated by the externally applied voltage and enhances or attenuates the local field. The resulting total electric field is dominated by a striation pattern, too. As a consequence, bulk electrons gain high energies only in narrow regions, whereas the electric field caused by the space charges greatly attenuates the drift electric field between these regions. Note that the electron density is very small in the plasma bulk due to the high electronegativity, so that the electrons hardly affect the space charge distribution, i.e., although the electrons can react much more quickly to the RF electric field, they cannot compensate the space charge generated by the ion motion. These electrons then cause excitation, dissociative attachment, and ionization in subsequent collisions with the neutral gas. The respective rates exhibit local maxima at the edges of the regions of high electric field, thereby contributing to the development of striations in the light emission (cf. Fig.~\ref{fig:simProes}) and the ion densities (Fig.~\ref{dens}(a)).

The spatial modulation of the electric field in the bulk does not vanish on time average. Instead, there are alternating regions of negative and positive field. The positions of the local maxima of the F$^-$ and CF$_3^+$ ion densities correspond to the positions of $E\approx0$ with a strongly negative field gradient, right of an adjacent local maximum of the mean electric field, as shown in Fig. \ref{fig:xts}(d). (Note that this figure shows the mean field in the bulk region only; the field strength strongly increases in the sheath regions.) The ion velocity is composed of an oscillatory part within the RF period and a steady drift part. The latter is proportional to the local mean field due to the high pressure. Hence, the drift motion in the mean field ensures that the local maxima of the ion densities remain stable, since it ``focuses'' the positive ions to these positions. The positive ions exhibit the largest density and dominate the heavy particle dynamics for reasons of quasineutrality, i.e., the negative ions are bound to the positive ion motion on time scales beyond the RF period.

\begin{figure*}[h!t]
\begin{center}
\includegraphics[width=0.99\textwidth]{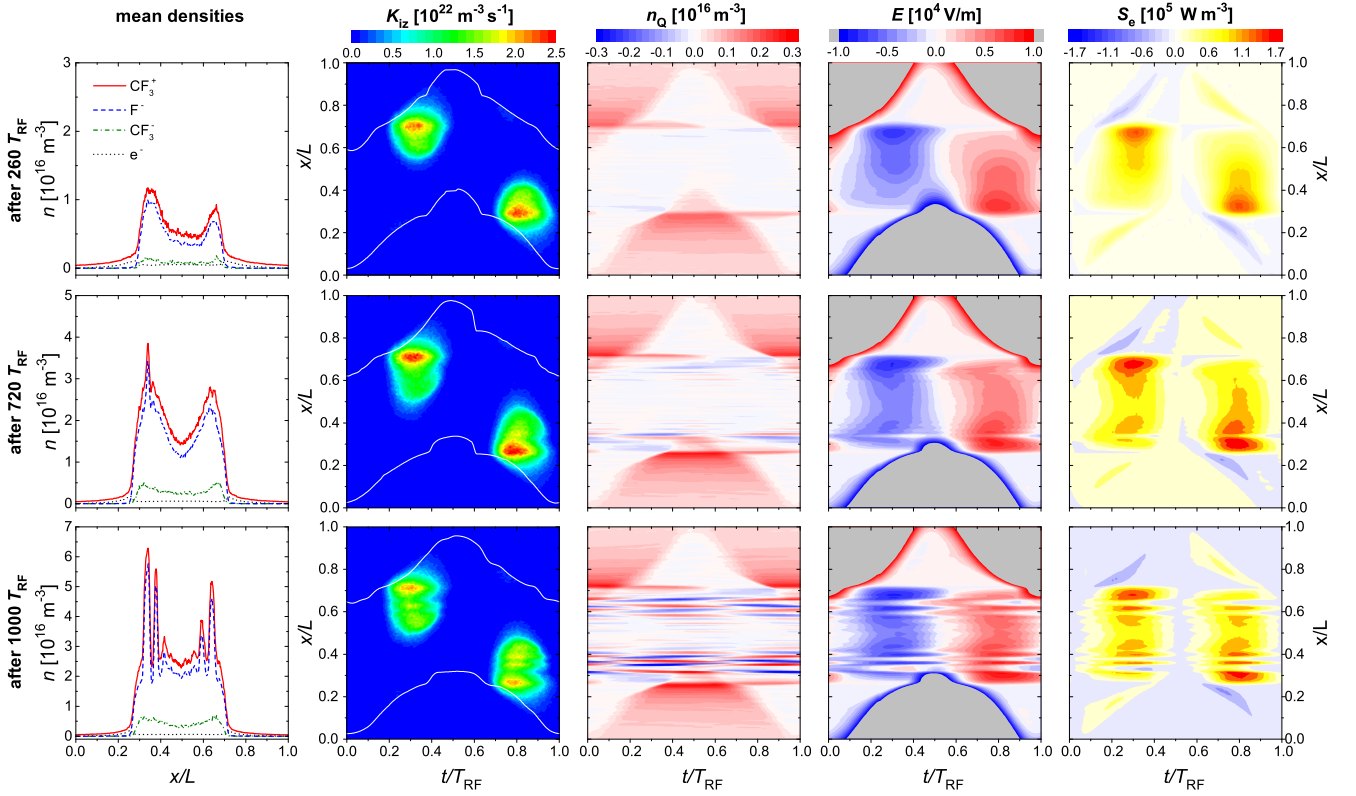}
\caption{(a) Density profiles and spatio-temporal plots of the (b) ionization rate (with sheath edges indicated by white lines), $K_{iz}$, (c) the charge density, $n_Q$, (d) the electric field, $E$, and (e) the power absorption rate by electrons, $S_e$, obtained from the PIC/MCC simulation at different times (after 260, 720, and 1000 RF periods) during the convergence process. Videos visualizing the evolution of these quantities are provided online \cite{striations_video}. The data are integrated for 20 RF periods.} 
\label{fig:conv}
\end{center}
\end{figure*}

In order to gain insight into the {\it formation mechanisms} of the striations, we track all relevant quantities in the simulation, following the seeding of initial electrons in the neutral gas filled electrode gap.
Figure \ref{fig:conv} summarizes the relevant distributions of the densities, ionization rate, space charge densities, electric field, and electron power absorption at different times after the initialization of the PIC/MCC code. At early times (i.e., 260 RF periods, first row), no striations are present. The spatio-temporal ionization rate exhibits two maxima close to the sheath edges at the times of collapsing sheaths. This is because electrons mainly gain energy from the interaction with the ambipolar field that develops due to the strong gradient in the local electron density \cite{ambipolar,WangAmbipolar}. Therefore, the electric field is  high in the regions $x/L \approx $ 0.3 and 0.7 at the times of collapsing sheaths, $t/T_{RF} \approx $ 0.3 and 0.8, respectively. As a consequence, power is efficiently dissipated to the electrons so that positive and negative ions are generated there. Consequently, the ion densities increase until the local ion densities become so large, that the ion plasma frequency is comparable to the applied frequency (see second row of Fig. \ref{fig:conv}, after 720 RF periods).


In such an ion-ion plasma, the plasma frequency, that defines the time scale on which the ions can react, is lower for a given ion species compared to an electron-ion plasma. This can be shown based on a homogeneous, pure ion-ion plasma model. The momentum balance equations 
\begin{equation}
m_{\pm} n_{\pm} \frac{\partial^2 x_{\pm}}{\partial t^2}= q_{\pm} n_{\pm} E - m_{\pm} n_{\pm} \nu \frac{\partial x_{\pm}}{\partial t} \nonumber
\end{equation}
of positive (index "+", $q_{+}=e$) and negative (index "$-$", $q_{-}=-e$) ions with masses $m_{\pm}$ and densities $n_{\pm}$ within an externally applied electric field $E_0 \cos(\omega_{rf} t)$ lead to two coupled differential equations
\begin{align}
 \frac{d^2 x_{-}}{d t^2} + \nu \frac{d x_{-}}{d t} - \omega_{-}^2 (x_{+}-x_{-}) &= -\frac{e}{m_{-}} E_0 \cos(\omega t), \nonumber\\
 \frac{d^2 x_{+}}{d t^2} + \nu \frac{d x_{+}}{d t} + \omega_{+}^2 (x_{+}-x_{-}) &= \frac{e}{m_{+}} E_0 \cos(\omega t), \nonumber
\end{align}
that describe the ion motion (i.e., the displacement $x_{\pm}$) of the positive and negative ions. Here, $\omega_{\pm}=\sqrt{e^2 n / \epsilon_0 m_{\pm}}$. The last term on the left hand side is caused by the development of space charges. Such a system will be in resonance with the driving oscillation and, hence, large oscillation amplitudes will occur if $\omega_{rf}^2 = (\omega_{-}^2 + \omega_{+}^2)$. Therefore, the "plasma frequency" based on the coupled motion of both ion species is lower compared with the plasma frequency of electrons in electropositive plasmas (where $\omega_{rf}^2=\omega_{\rm pe}^2$). This means that plasma densities of about $2.2 \cdot 10^{16}$ m$^{-3}$ are sufficient to match the resonance condition above for a CF$_3^{+}$--F$^{-}$ plasma driven by $\omega_{rf}= 2 \pi \times 8\cdot 10^6$ s$^{-1}$. The eigenfrequency depends on the ion masses and the plasma density. Hence, in general a resonance will occur also for heavy ions and a high driving frequency, if the density is sufficiently high.

Further, we note that the spatial profiles and dynamics are similar to that in the formation process of double layers \cite{eneg3,eneg4,eneg5,eneg6,enegA,Sheridan1,Sheridan2,Sheridan3}, 
where the total densities in the bulk remain about the same. In the present case, however, the ion densities in the plasma bulk rise further, so that the motion of the ions causes a significant space charge and affects the electric field. On the one hand, the field is attenuated at the position of the density peak. On the other hand, the field is amplified adjacent to the density peak, leading to enhanced electron power absorption. Hence, the ionization rate is increased at a small distance from the initial density striations. Subsequently, the density is further enhanced and a new pair of striations is formed by the same mechanism as described above (third row of figure \ref{fig:conv}, after 1000 RF periods). Thus, the bulk region is filled with striations layer by layer, starting at the plasma sheath edges and proceeding towards the discharge center. Finally, the entire bulk region is filled with striations in the steady state, as shown in the analysis of the fully converged simulation results above. We note that the formation of the striations at both electrodes is not completely symmetric (see Fig. \ref{fig:conv}, left column). This is caused by statistical effects, i.e. the threshold ion density is reached slightly earlier adjacent to the powered electrode. As a consequence of this asymmetry a DC self-bias is generated during the convergence of the simulation, which decreases and finally vanishes, when the simulation is fully converged.

In conclusion, we studied the formation and sustainment of layered, striation-like structures in electronegative capacitive discharges. The development of these structures, which are observed in the spatio-temporal plasma emission profile for the first time here, is linked to the periodic acceleration of ions by the RF electric field. The ion motion generates local space charges that cause a spatial modulation of the electric field. This strongly affects the electron heating rate and, thereby, the ionization rate. Thus, once the ion number density is large enough (i.e., exceeds a threshold density) at a certain position within the plasma bulk to allow the ions to react to the RF electric field and, consequently, to form space charges, which affect the total electric field, the ionization rate is increased around this position. The formation of the striations is, therefore, based on the resonance of the ion-ion plasma. This novel effect is not only of fundamental importance, but will affect the ion flux in electronegative CCRF plasmas used for surface processing applications, which are typically operated under similar conditions. 
The present study lays the foundation for a broad research of this type of striations. The striation structure is expected to be very sensitive to the ion species (i.e., the ion mass in electronegative gas plasmas such as O$_2$, SF$_6$, C$_4$F$_8$), collision frequency, and driving frequency. (For instance, the striations will vanish at high driving frequencies, if the resonance condition is not met.)  Such a detailed experimental investigation is in progress.



This work has been financially supported by the Hungarian Scientific Research Fund, via Grant OTKA K105476 and by the National Natural Science Foundation of China (NSFC) (Grant Nos. 11335004 and 11405018).


\end{document}